\documentclass[useAMS,usenatbib]{mn2e}
\usepackage{aas_macros}
\usepackage{graphicx}
\usepackage{color}
\usepackage{subfigure}
\usepackage{url}
\usepackage{float}

\newcommand{\gadg}{\textsc{gadget }}
\newcommand{\gadgt}{\textsc{gadget-$3$}}

\title[Softening in galaxy simulations]{The effect of softening on dynamical simulations of galaxies}
\author[F.~Iannuzzi and E.~Athanassoula]{Francesca Iannuzzi\thanks{E-mail:francesca.iannuzzi@oamp.fr} and E.~Athanassoula\\Aix Marseille Universit\'e, CNRS, LAM (Laboratoire d'Astrophysique de Marseille) UMR 7326, 13388, Marseille, France}
\begin{document}

\date{Accepted 2013 August 29.  Received 2013 August 26; in original form 2013 May 8}

\pagerange{\pageref{firstpage}--\pageref{lastpage}} \pubyear{2013}

\maketitle

\label{firstpage}

\begin{abstract}
Dynamical simulations are a fundamental tool for studying the secular evolution of disc galaxies. Even at their maximum resolution, they still follow a limited number of particles and typically resolve scales of the order of a few tens of parsecs. Generally, the spatial resolution is defined by (some multiple of) the softening length, whose value is set as a compromise between the desired resolution and the need for limiting small-scale noise. Several works have studied the question whether a softening scale fixed in space and time provides a good enough modelling of an astrophysical system. Here we address this question within the context of dynamical simulations and disc instabilities. We first follow the evolution of a galaxy-like object in isolation and then set up a simulation of an idealised merger event. Alongside a run using the standard fixed-softening approach, we performed simulations where the softening lengths were let to vary from particle to particle according to the evolution of the local 
density field in space and time. Even though during the most violent phases of the merging the fixed-softening simulation tends to underestimate the resulting matter densities, as far as the evolution of the disc component is concerned we found no significant differences among the runs. We conclude that using an appropriate fixed softening scale is a safe approach to the problem of modelling an N-body, non-cosmological disc galaxy system. 
\end{abstract}

\begin{keywords}
methods: numerical -- galaxies: evolution -- galaxies: kinematics and dynamics.
\end{keywords}

\section{Introduction}
\label{sec:intro}
Numerical simulations are a tool of paramount importance in the study of galaxy formation and evolution.
Even if one is interested in assessing the effect of gravity only, a fully consistent picture of the assembly of matter and its subsequent evolution is beyond the reach of analytical techniques and can only be provided by so-called ``N-body'' simulations.
Notwithstanding the steady progress that numerical modelling has experienced in the past forty years, it is still non-trivial to account for all the desired and relevant processes at once. Faithfully simulating the growth of a galaxy within a cosmological context may result in adopting an insufficient resolution to address the issue of its dynamical evolution. One therefore still has to tailor, somehow, the features of the simulation to the specifics of the problem.\\
The numerical modelling of an idealised and isolated galaxy-like system trades a realistic treatment of environmental effects for an enhanced resolution. Such simulations -- also termed as ``dynamical'' -- provide an ad-hoc laboratory to study the development of dynamical instabilities and the internally-driven, ``secular'' evolution of disc galaxies \citep[see][for a review]{athanassoula12}. Starting from simulations with a handful of particles such as those run in the 1970s, a number of works have since crucially contributed to our current understanding of disc dynamics. In particular, the phenomena of bar formation and evolution, along with its dependence on the properties of the dark-matter halo and the disc itself has been thoroughly studied by numerical means \citep[][among others]{miller70,ostriker73,athanassoula86,combes90,athanassoula96, debattista00,athanassoula02,athanassoula03,martinezvalpuesta06,dubinski09,athanassoula13, saha13}.\\
Such simulations generally involve the modelling of up to a few million particles, a  number several orders of magnitude lower than the actual number of stars building up a typical disc. This under-representation of the system's distribution function, along with the inevitable discreteness of the time integration, result in the need to soften the interaction between particles on small scales. In practical terms, this means substituting the inverse-square law with a milder, non-diverging interaction. This helps in limiting the noise due to a spurious collisional behaviour on small scales, whilst keeping the computational time down to reasonable values. Confining the noise due to long-range encounters remains a matter of increasing the total number of particles used, but this source of collisionality may not be as important in a disc-like configuration \citep{rybicki72,sellwood13}.\\
A commonly adopted approach to softening is to identify one reasonable spatial scale below which to alter the nature of the gravitational force and to apply it as such to all interactions throughout the simulation. How the scale is chosen depends on the features of the system and is subject to some level of arbitrariness \citep{merritt96, melott97, romeo98, athanassoula00, knebe00, dehnen01, power03, rodionov05, zhan06}. It is normally taken to be some fraction of the mean interparticle separation and in the simulations under consideration here it can reach some tens of parsecs. This scale represents a strict resolution limit below which any information is physically meaningless. \\
Whether or not having the softening length fixed in space and time is a reasonable approach to modelling a system has been subject of debate \citep{bate97,dehnen01, binney02, nelson06, romeo08, bk09}. In a system where the density field is subject to wild variations in space and/or time, having a fixed resolution may result in excess noise in under-dense regions while providing a completely biassed representation of the over-dense ones. Whether this severely impacts the outcomes of dynamical simulations is the subject of our paper. \\
In what follows, we are going to show the results of two sets of simulations where identical initial conditions have been evolved with different softening approaches. Alongside the standard run with fixed softening, we have performed a few other simulations where the softening lengths were individually defined for each particle and allowed to vary with time. The algorithm behind this technique has been presented by \cite{pm07} and it defines the softening scale as the radius of the sphere containing a given number of neighbours $N_{ngbs}$ -- exactly the same definition as for smoothing lengths in SPH. This way the gravitational resolution adapts to the features of the density field, varying proportionally to the local mean interparticle separation. The side effect of introducing such a flexible resolution is in that the equation of motion needs to be changed accordingly. Indeed, the application of the Euler-Lagrange equation to the modified Lagrangian of the system brings about an extra-term besides the 
standard, Newtonian force law. A correct implementation of this quantity -- which we will hereafter refer to as the ``correction term'' -- is crucial in order to guarantee energy conservation throughout the simulation. \\
\cite{iannuzzi11} have implemented the adaptive-softening algorithm in the simulation code \gadgt~ and applied it to a number of test cases and cosmological simulations. 
In problems with an analytical solution, they showed that adaptive softening provides a more faithful representation of the force field and density distribution without any fine-tuning of the parameter $N_{ngbs}$. Conversely, the results were found to be much more sensitive to the choice of the softening scale when this was taken to be fixed. When investigating the effect on cosmological simulations, it was found that, given the same total number of sampling particles, using the standard prescription for fixed softening caused an underestimation of clustering in overdense regions. Overall, it was concluded that letting the softening length vary was providing optimal softening with little dependence on the adopted number of neighbours and that the consequences were particularly evident in cases with a markedly inhomogeneous density field. For the latest application of the method to a dark matter, ``zoom-in'' cosmological simulation of a galactic halo, see Sec.~5 of \cite{kim13}.\\

In this work we want to understand whether the results on the dynamical evolution of disc galaxies, as obtained from dedicated simulations, are affected by the adoption of a fixed resolution. We will now move on to introduce the features of the simulation sets we have performed for this study, whose results are presented in Sec.~\ref{cusp} and \ref{merger}. We will then summarise the findings and draw our conclusion in Sec.~\ref{concl}.

\section[]{Simulations and results}
\label{sec:simsres}
The first simulation set performed in this study features an idealised and isolated galaxy system consisting of a stellar disc embedded in a dark-matter halo. We have initially considered two set-ups, differing in the properties of the host halo: in one case this consisted of a ``core'', meaning that the central density distribution is flat, while in the second it represented a so-called ``cusp'', whereby the central density behaves like a power-law with index $-1$. Since the conclusion regarding the effect of softening on the final properties of the system is very similar in the two cases, we have decided to show the results for the ``cusp'' simulation-set only. The reason for this choice is that the latter case is somewhat more interesting from a numerical point of view. Indeed, an increasing density towards the centre represents a more challenging scenario for an N-body code, especially from the point of view of identifying an appropriate resolution scale.\\
The second simulation set that we are going to discuss consists of an idealised merger. A satellite shaped as a Hernquist sphere \citep{hernquist90} is let to collide with an equilibrium disc+halo system which is initially identical to the the ``core'' system just mentioned. The trajectory was chosen to be perpendicular to the disc plane and passing through its centre. As a consequence of the geometrical configuration of the encounter, a ring develops in the disc component. Eventually, a bar forms as the system reaches an equilibrium whereby the remnant of the satellite settles around the centre of the galaxy-like object.\\
The initial conditions for the two disc+halo systems under consideration were generated with the iterative method of \cite*{rodionov09}. The Hernquist sphere was set up with the local Maxwellian approximation and its equilibrium properties carefully monitored before it was incorporated in the initial conditions of the merger. Both the isolated galaxy and merger system were evolved for ten gigayears with the code \gadg \citep{springel01b, springel05} in its latest version \gadgt. Within each simulation set, both the makefile options and the adopted parameters are identical, except for those concerning the choice of gravitational softening. \\
In the model with fixed gravitational softening (hereafter referred to as ``Fix'') the choice for the plummer-equivalent softening $\epsilon$ was set to $\epsilon = 0.05\;\rm{kpc}$. This corresponds to the reference value adopted in previous works using the same initial conditions \citep*[as, for example, in][]{athanassoula13}. As we show in Appendix \ref{app}, we also ran both the ``core'' and ``cusp'' models in isolation with $\epsilon = 0.025\;\rm{kpc}$  and $\epsilon = 0.1\;\rm{kpc}$, finding no differences in the macroscopic quantities of interest here. \\
When using the adaptive algorithm, the parameter to set is the number of neighbours $N_{ngbs}$. For a particle $i$, this quantity is related to the individual softening $h_i$ via the following relation:
\begin{equation}\label{ngbs_h}
\frac{4\pi}{3} h_i^3 \sum_{j=0}^{N} W_{ij}(h_i) = N_{ngbs},
\end{equation}
where $W$ is the adopted smoothing kernel (here a cubic spline) and the sum is over all the $N$ particles found within a distance $h_i$ from particle $i$. Note that, in general, $N \neq N_{ngbs}$ and the relation between the two quantities depends on the particle distribution and the shape of the kernel. Indeed, a choice of $N_{ngbs} = 60$ results in individual softenings of sizes around twice the local mean interparticle separation only \citep[see][]{pm07}, which is less than what expected if the softening sphere were really to comprise sixty particles. In all simulations presented in this paper and as far as adaptive softening is concerned, the neighbour search does not distinguish between particles belonging to different components (e.g. ``halo'' vs. ``disc'' particles). We have performed two runs with the fiducial choice of $N_{ngbs} = 60 \pm 0.3$, one with correction of the equation of motion (hereafter referred to as ``A60'') and one without (hereafter referred to as ``noC60''). In the ``cusp'' 
simulation, we also performed one run with $N_{ngbs} = 30 \pm 0.3$ and no correction to the equation of motion (hereafter referred to as ``noC30''). Adopting thirty neighbours or less has proven from our previous studies not to be safe in terms of obtaining a robust evaluation of the correction term \citep[see, e.g.,][]{iannuzzi11}, therefore we decided not to run the fully conservative algorithm in this case. We recall that the correction term is an additional expression that should be accounted for in the equation of motion when adaptive softening is used, in order for the total energy to be conserved. Hence, for the simulations marked as ``noC'' conservation of energy is not strictly guaranteed. However, even though the temporal variation of individual softening lengths was 
indeed causing fluctuations or drifts in the total energy budget of the test cases performed in \cite{iannuzzi11}, the lack of the appropriate correction term in the equation of motion does not seem to have as much of an influence in the simulations presented in this case (where energy is conserved at a $\approx 0.2\%$ level, unless stated otherwise). This is presumably due to the larger number of particles involved here, while $N_{ngbs}$ remains effectively unchanged. The smoothed contribution to each particle's potential is therefore considerably smaller than for test cases with a handful of particles, leading to a larger insensitivity of the total potential to temporal variations in individual softenings.\\
Given such good energy-conservation properties of the noC simulations, we decided to show also the results from these runs alongside those from the fully-consistent adaptive algorithm. This allows us to push the number of neighbours down to thirty and explore the effect of much smaller adaptive softenings, without having to worry about evaluating the correction term accurately.\\
In the following subsections we will discuss the results from each of the simulation set, focussing mainly on the disc and related bar properties.
\begin{figure*}
\includegraphics[width=178mm]{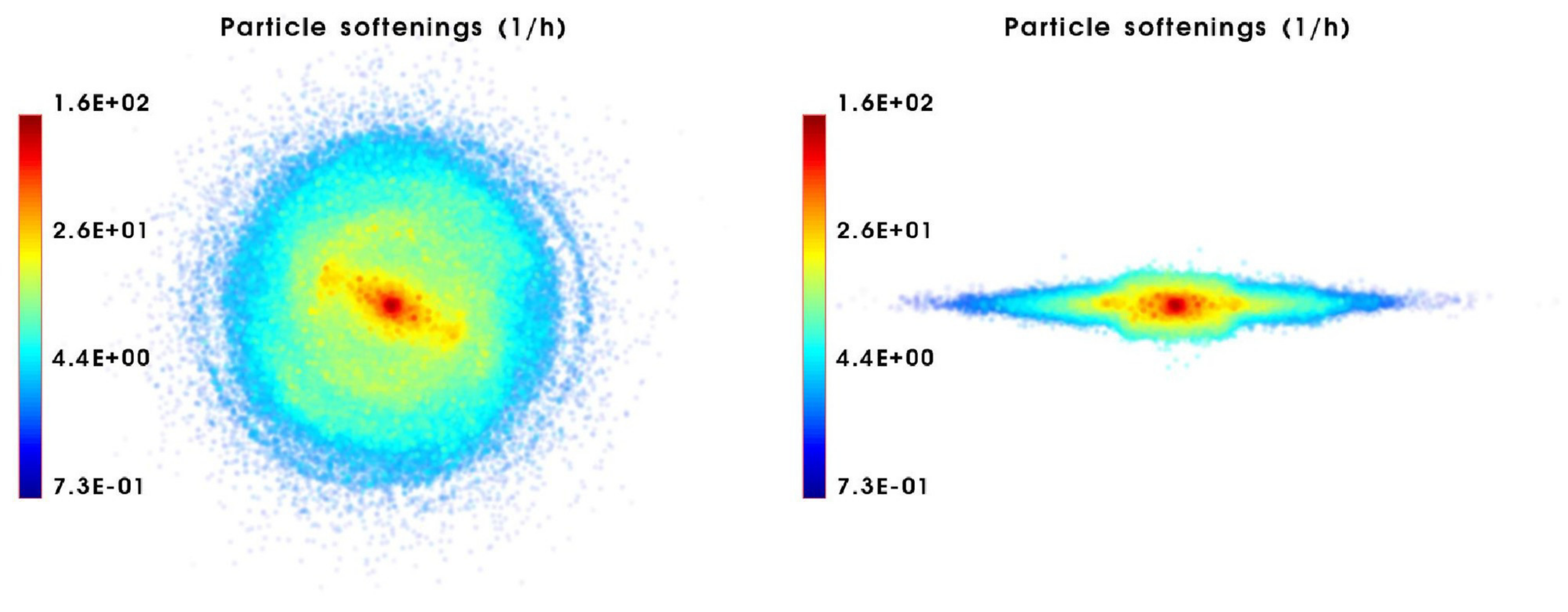}
 \caption{Face-on and edge-on appearance of the disc component for the ``cusp'' simulation noC30 at $t=10\;\rm{Gyr}$. Particles are coloured according to the inverse of their associated softening length. The values in terms of $h$ range from $0.006\;\rm{kpc}$ (reddest points) to $1.37\;\rm{kpc}$ (bluest regions). As a reference, the fixed-softening value $\epsilon = 0.05 \;\rm{kpc}$ corresponds to the yellow-green points marked $20$ in the colour bar.}
  \label{agsh_combined}
\end{figure*}
\begin{figure}
\includegraphics[width=84mm]{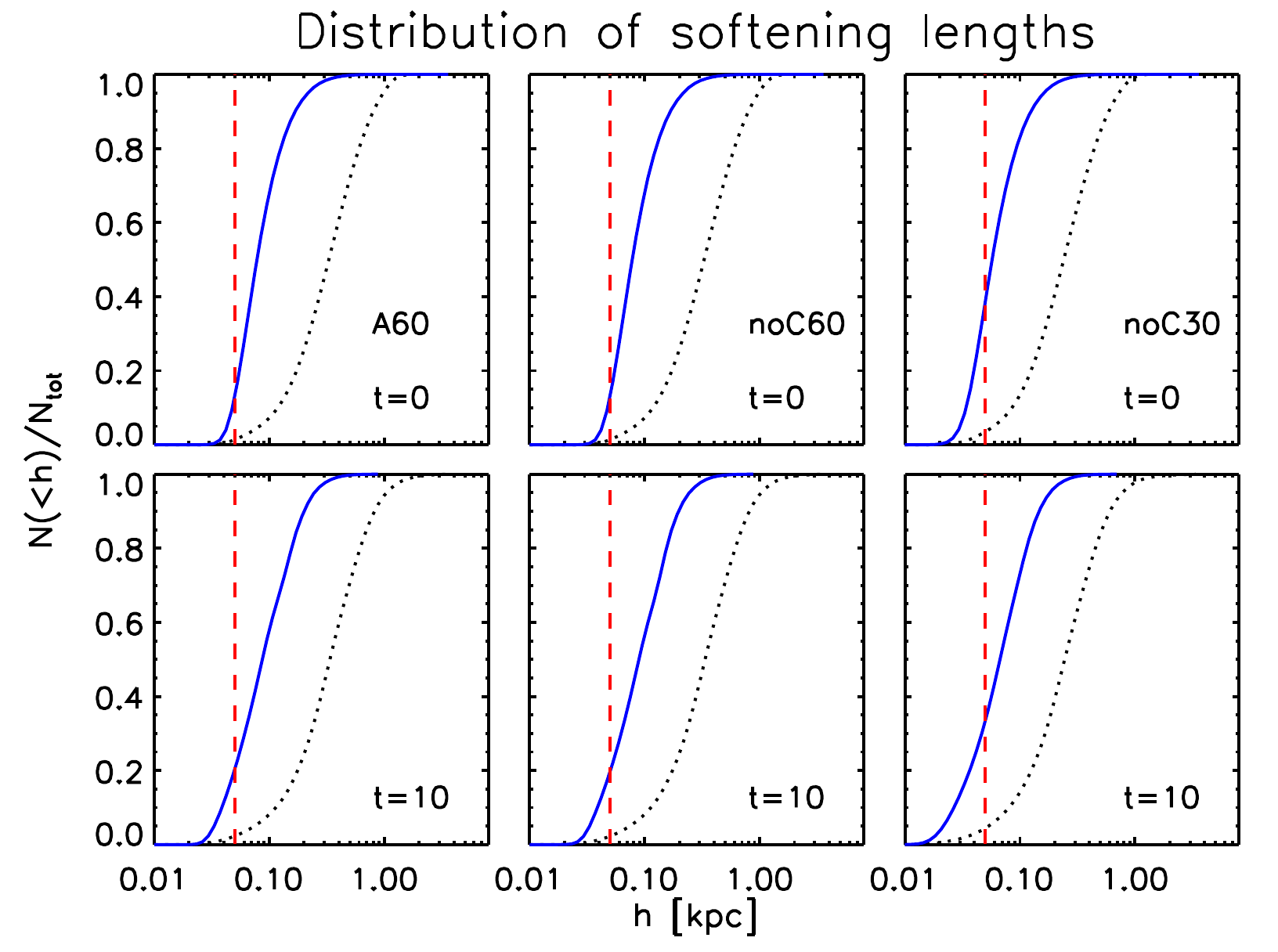}
 \caption{Cumulative distribution of softening lengths for the disc (blue, solid line) and the halo (black, dotted line) components in the ``cusp'' simulations. The vertical line marks the value of $\epsilon$ adopted in the fiducial simulation with fixed softening.
 Results are shown at the beginning ($t=0$, top row) and at the end ($t=10\;\rm{Gyr}$, bottom row) of the simulations. The three columns correspond 
 to the three adaptive-softening approaches under consideration and are named according to the convention introduced in Sec.~\ref{sec:simsres}. }
  \label{nfw_soft}
\end{figure}
\subsection{Cusped halo}\label{cusp}
\begin{figure*}
\includegraphics[]{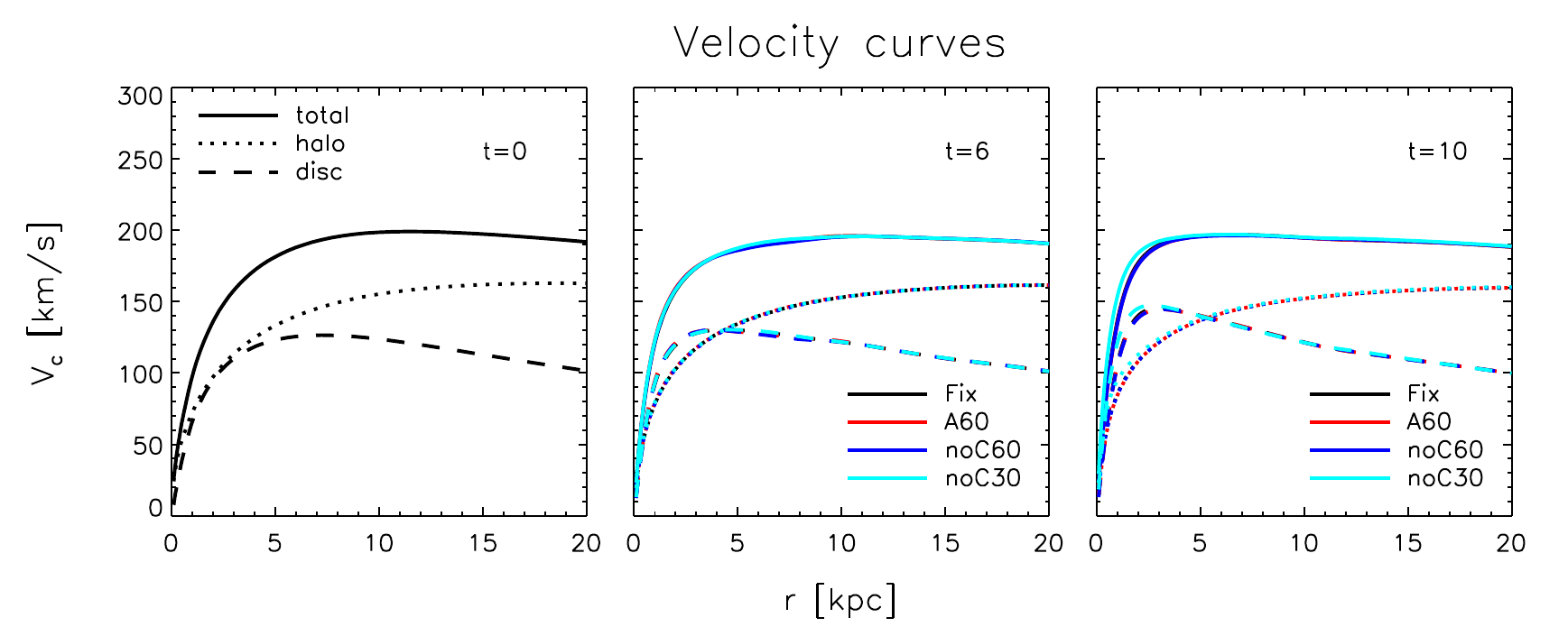}
 \caption{Circular velocity curves for the ``cusp'' simulations. The dashed and dotted lines give the contribution of the disc and halo, respectively, while the solid curve shows the results for both components together. The leftmost plot shows the velocity curves at the initial conditions, 
 which are common to all simulations. The central and rightmost plots give the results at $t=6\;\rm{Gyr}$ and  $t=10\;\rm{Gyr}$, respectively. Overplotted in different colours are the outcomes of the four softening approaches adopted, introduced in Sec.~\ref{sec:simsres}. }
  \label{nfw_vcurve}
\end{figure*}

\begin{figure*}
\includegraphics{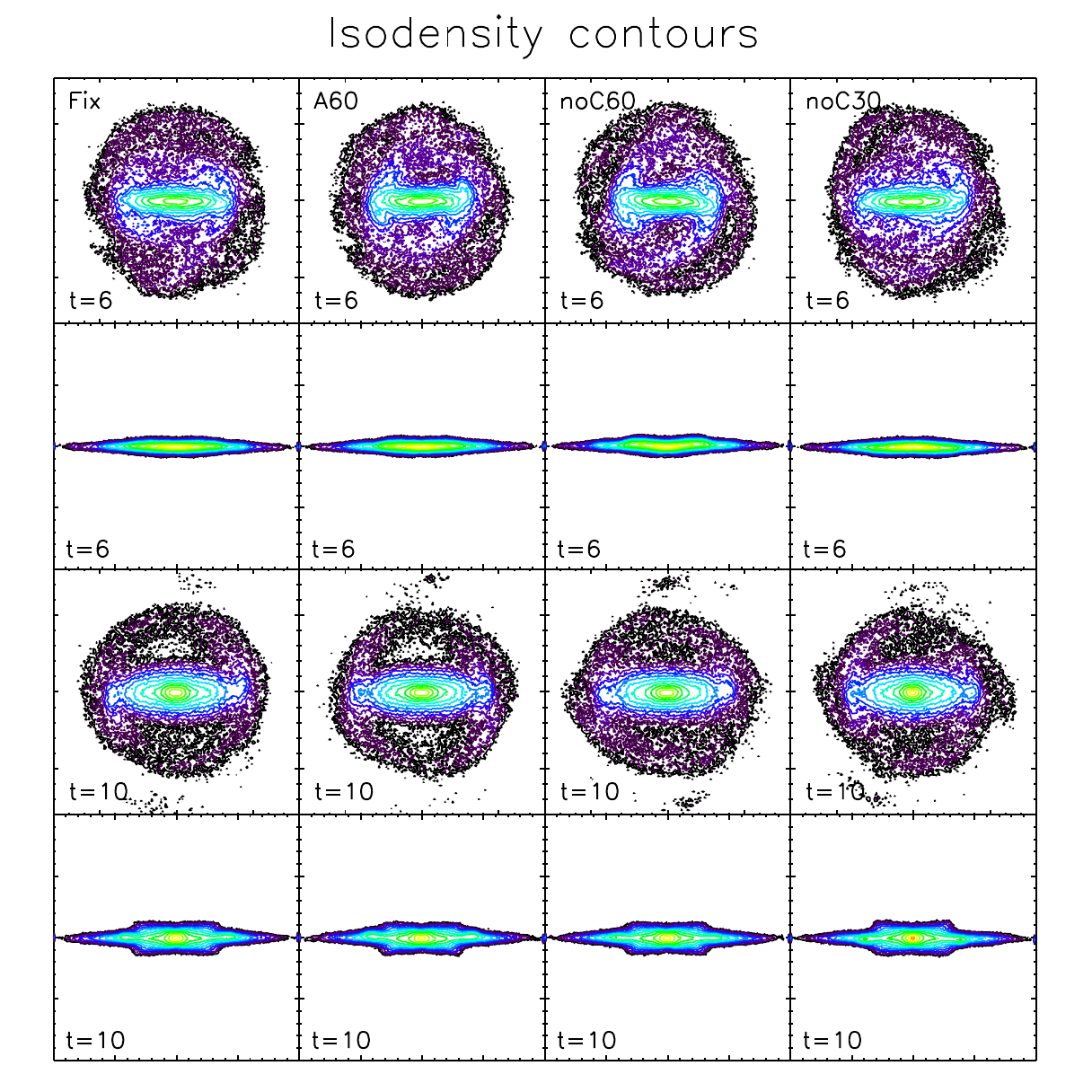}
 \caption{Isodensity contours for the disc component in the ``cusp'' simulations. The surface density is computed by projecting the particles on to the plane of the disc (face-on view, first and third row) and on to the perpendicular plane containing the major axis of the bar (side-on view, second and fourth row). Results are shown at $t=6\;\rm{Gyr}$, approximately at the end of the initial growth-phase for the bar, and at $t=10\;\rm{Gyr}$, marking the end of the simulation. Each column corresponds to a different softening approach, as explained in Sec.~\ref{sec:simsres}. The minor-tick interval marks everywhere a distance of $3\;\rm{kpc}$.}
  \label{nfw_isod}
\end{figure*}

\begin{figure*}
\includegraphics{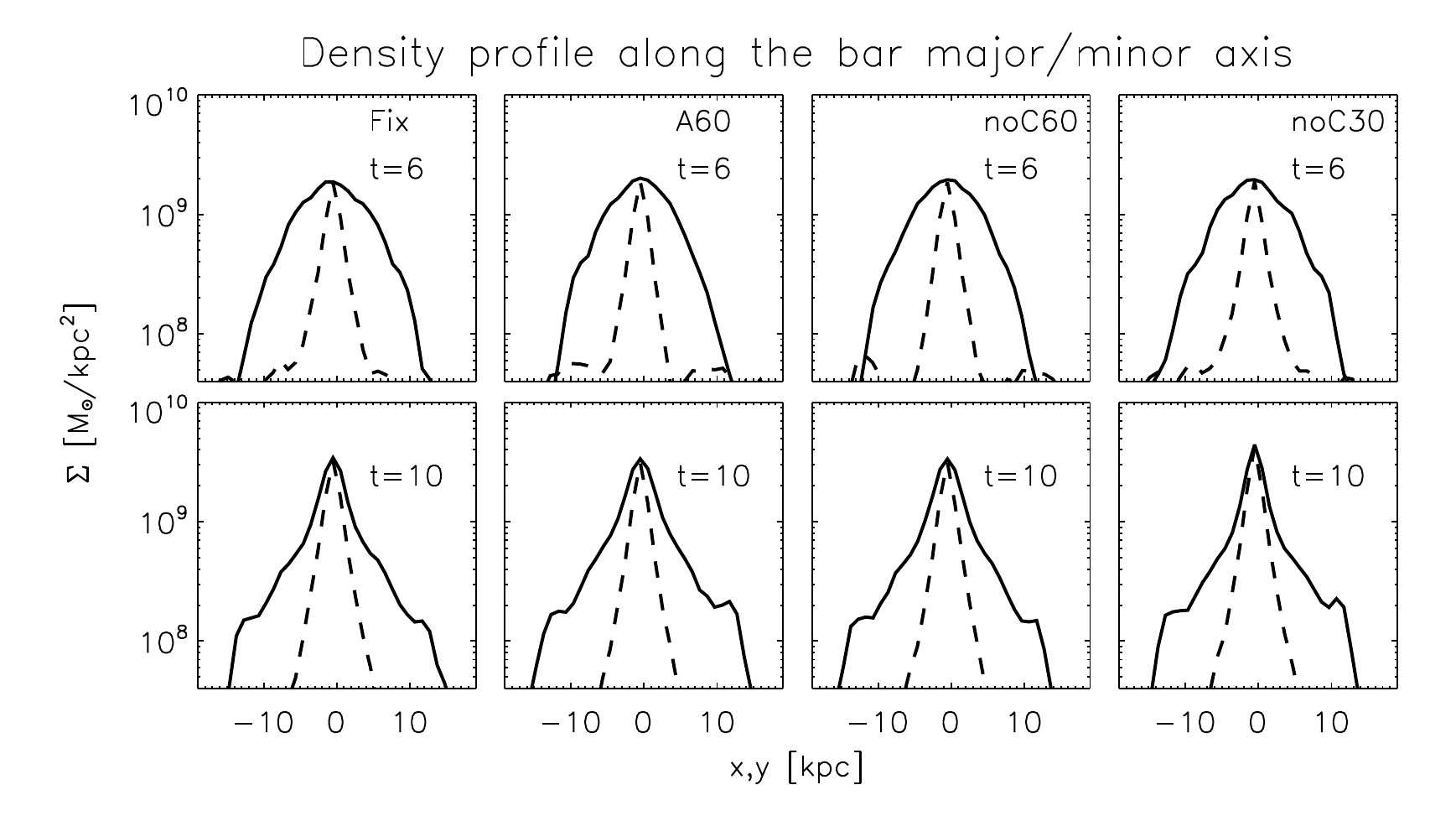}
 \caption{Projected density profiles along the bar major and minor axes (solid and dashed lines, respectively) for the ``cusp'' simulations. The surface density is computed by projecting the particles on to the plane of the disc and by selecting those within $0.5\;\rm{kpc}$ from the axis under consideration. Results are shown at $t=6\;\rm{Gyr}$, approximately at the end of the initial growth-phase for the bar, and at $t=10\;\rm{Gyr}$, marking the end of the simulation. Each column corresponds to a different softening approach, as explained in Sec.~\ref{sec:simsres}.}
  \label{nfw_dens_prof}
\end{figure*}

\begin{figure*}
\includegraphics[]{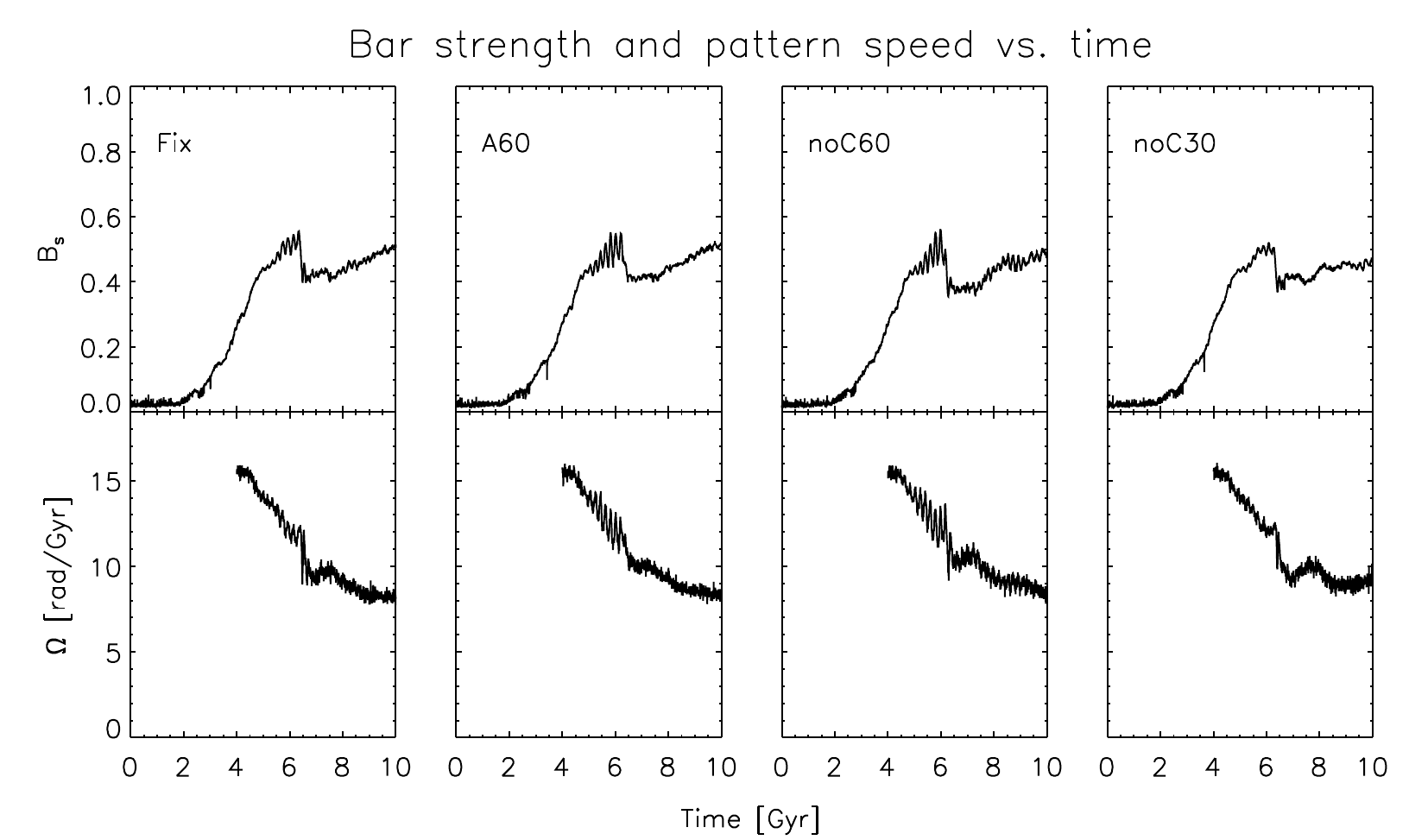}
  \caption{Bar strength and pattern speed as a function of time for the ``cusp'' simulations. 
  Here, the bar strength (upper panel) is defined as the relative amplitude of the $m=2$ Fourier components of the disc (see Eq.~\ref{bs} and the explanation in Sec.~\ref{cusp}). The pattern speed (bottom panel) measures the changes in the phase angle of the bar with time. Its evolution is shown starting from $t=4\;\rm{Gyr}$, due to a very noisy behaviour at earlier times. This is related to the bar not being fully formed yet and, consequently, its phase angle being very poorly defined. Each plot shows the results of a different softening approach, as explained in Sec.~\ref{sec:simsres}. }
  \label{nfw_bs_and_ps}
\end{figure*}

In terms of global features, the system consists of a  $5 \times 10^{10}\;\rm{M_{\odot}}$  disc embedded in a $2.4 \times 10^{11}\;\rm{M_{\odot}}$ spherical halo. The disc can be thought of as an old stellar component and it is modelled as a pure N-body system. It has an exponential density distribution characterised by a scale length $h=3 \;\rm{kpc}$ and a scale height $ z_0=0.6 \;\rm{kpc}$. The density distribution of the halo is given by an NFW profile \citep*{nfw} with scale radius $r_s = 11\;\rm{kpc}$ and concentration $c_{200}=12$, truncated at a radius $r_{max} = 60\;\rm{kpc}$. The total number of particles employed to represent the system is $2.8$ million, of which $2$ million go into the halo component and the remaining $800\;000$ into the disc. Correspondingly, the particle masses are $m_{halo} = 1.2 \times 10^5 \;\rm{M_{\odot}}$ and  $m_{disc} = 6.25 \times 10^4\;\rm{M_{\odot}}$ for the halo and disc components, respectively.\\
A visual representation of the adaptivity of the softening lengths is provided by Fig.~\ref{agsh_combined}. The face-on and edge-on view of the disc component for the noC30 simulation is shown at $t=10\;\rm{Gyr}$, i.e. the end of the simulation. Each particle is coloured based on the value of its associated softening length\footnote{Actually, particles are coloured on the basis of the inverse of their softening. This choice has been made solely to improve the quality of the visualisation.}, and the extent to which the variation in softenings reflects the variation in the density field is evident. For comparison, the fixed-softening value $\epsilon$ corresponds to a value of $20$ in the colour scale and it is associated to particles at the border or just outside the bar. This means that the adaptive algorithm naturally tends to assign smaller softenings than $\epsilon$ within the bar region and larger softenings in the outskirts.\\
A more quantitative description of the differences between the four runs in terms of assigned softenings is provided by Fig.~\ref{nfw_soft}. The cumulative distribution of softening lengths is shown at the beginning (top row) and at the end (bottom row) of the simulation for the different cases. The results for the disc and halo component are shown separately, by the blue-solid and black-dotted lines respectively. The vertical, dashed line marks the value of fixed softening used in the Fix simulation, i.e.  $\epsilon = 0.05 \;\rm{kpc}$. The two simulations with 60 neighbours have very similar softening distributions. For the disc component, the percentage of particles with associated softening smaller than $\epsilon$ grows from $15\%$ at $t=0$ to just above $20\%$ at $t=10\;\rm{Gyr}$. For the halo component this percentage also increases with time, but hardly ever exceeds few per cents. As intuitively expected, the simulation with thirty neighbours induces much smaller softenings. Indeed, the 
disc 
particles with softening smaller than $\epsilon$ are about $40\%$ of the total at all times and the percentage increases also for the halo component. The smallest registered softening length in the noC30 simulation is $0.006 \;\rm{kpc}$, approximately a factor of eight smaller than $\epsilon$. On the other hand, the largest recorded value is more than a factor 20 larger than  $\epsilon$, exceeding $1\;\rm{kpc}$.\\
In the analysis of the resulting matter distribution, we first look at the velocity curves. We evaluate them by measuring the amount of matter found within spheres of increasing radii and centred on the potential minimum of the galaxy.
Fig.~\ref{nfw_vcurve} shows the results for the different runs at three representative times during the evolution. The left panel corresponds to the initial conditions, identical for all the runs. The middle panel shows the result at $t= 6\;\rm{Gyr}$, marking the end of the initial growth-phase of the bar, while the right panel displays the curves at the end of the simulation. Regardless of the softening approach adopted, the curves for both the halo and disc component evolve very similarly. This means that the cumulative, spherically-averaged matter distribution in both components is practically identical in all runs. An exception to this may be seen in the late-time behaviour of the noC30 simulation, where matter seems to be more concentrated than in the other cases. Whether this behaviour is entirely benign, though, is not certain. It arises starting from $t\approx7\;\rm{Gyr}$ and in parallel with a slight drift in the total energy of the system, which results in the final value being $\approx 0.8\;\%$ 
lower 
than at the start. This less accurate behaviour of the simulations may 
play a role in generating this enhanced central concentration of matter.\\
A visual inspection of the disc component does not point to any significant difference between the runs either. Fig.~\ref{nfw_isod} shows the isodensity contours for the face-on and edge-on distribution of the disc particles, for all the simulations and at the two representative times  $t=6\;\rm{Gyr}$ and  $t=10\;\rm{Gyr}$. The results are definitely not identical, especially when looking at the edges of the bar and the outer regions, but in the context of the results obtained from more quantitative analysis we deduce that the differences are not significant. Also, the fact that we are showing the results at the same time does not mean that the discs should be in exactly the same evolutionary stage. In this sense, the differences in the asymmetries of the edge-on profile are not a concern. Similar conclusions apply when comparing the results at all other times during the simulation.\\
The projected density profiles along the bar's major and minor axis are shown in Fig.~\ref{nfw_dens_prof}. The absolute values of $\Sigma$ and its spatial evolution away from the centre are similar, regardless of the softening approach adopted. This holds for both the behaviour along the major and minor axes, marked by the solid and dashed line respectively, and at all times. We note that at late times, in coherence with what was previously shown in terms of velocity curves, the central density in the noC30 simulation peaks at a somewhat larger value than in the other runs.\\
Last, Fig.~\ref{nfw_bs_and_ps} shows the behaviour of the bar strength and pattern speed as a function of time. The global strength of the bar is here defined as
\begin{equation}\label{bs}
B_s = \frac{\sum_i \sqrt{a_2^2+b_2^2}}{\sum_i a_0},
\end{equation}
where $a_m$ and $b_m$ are the even and odd Fourier components of order $m$ and the sum extends over the annuli $i$ enclosed within the region where the local bar strength -- $\sqrt{a_2^2+b_2^2}/ a_0$ -- is above $50\%$ of its maximum. As shown by the curves in the first row, the importance of the bar grows with time until around $t=6\;\rm{Gyr}$, where it reaches a maximum and then rapidly decays as the buckling - i.e. the development of vertical asymmetries - occurs \citep[see, e.g.,][]{combes90,raha91,martinezvalpuesta06,athanassoula08}. Soon after, the so-called ``secular evolution'' phase, characterised by a slow and regular growth of the bar, takes over \citep[see][for a review]{athanassoula12}. The time of the buckling differs from one run to the other only by a few tens of million years. In general, both the initial-growth and secular-evolution phases show striking similarities in terms of absolute values for the bar strength and its temporal run. The exception may be the final stage of the secular 
evolution for the noC30 simulation, where the bar does not seem to 
grow as much as in the other cases. This can again be related to the excess in the central concentration of matter that occurs in the last few gigayears of evolution, which results in a suppression of the growth of the bar \citep{hasan90, shen04, athanassoula05}. We also note an oscillating behaviour of $B_s$ just before the buckling occurs. This is more evident in the noC60, A60 and Fix simulations, whereas noC30 shows a much milder feature. We remark that the results from  noC30 are fully reliable in this time range and up to around $t=7\;\rm{Gyr}$, as energy is conserved equally well as in the other simulations. By inspecting the time evolution of the Fourier components as a function of radius, as well as the time evolution of the particle distribution itself, we have identified the origin of this oscillatory behaviour in the transfer of small-mass blobs alongside the bar. This feature may well be excited in a manner dependent on the adopted softening -- as the striking similarities between A60 and noC60,
 along with the difference with respect to Fix and noC30, seem to suggest. It remains, however, a transitory feature and, in addition, by smoothing $B_s$ we would recover very similar results from all the runs. The second row of Fig.~\ref{nfw_bs_and_ps} shows the rotation speed of the bar. The results are shown starting from  $t=4\;\rm{Gyr}$, as noise dominates at earlier time due to the phase angle of the bar being poorly defined. The behaviour of the pattern speed is intimately related to the time evolution of the bar strength, a well-known result which can also be appreciated in these plots.

\subsection{An idealised merger}\label{merger}
\begin{figure*}
\includegraphics[width=178mm]{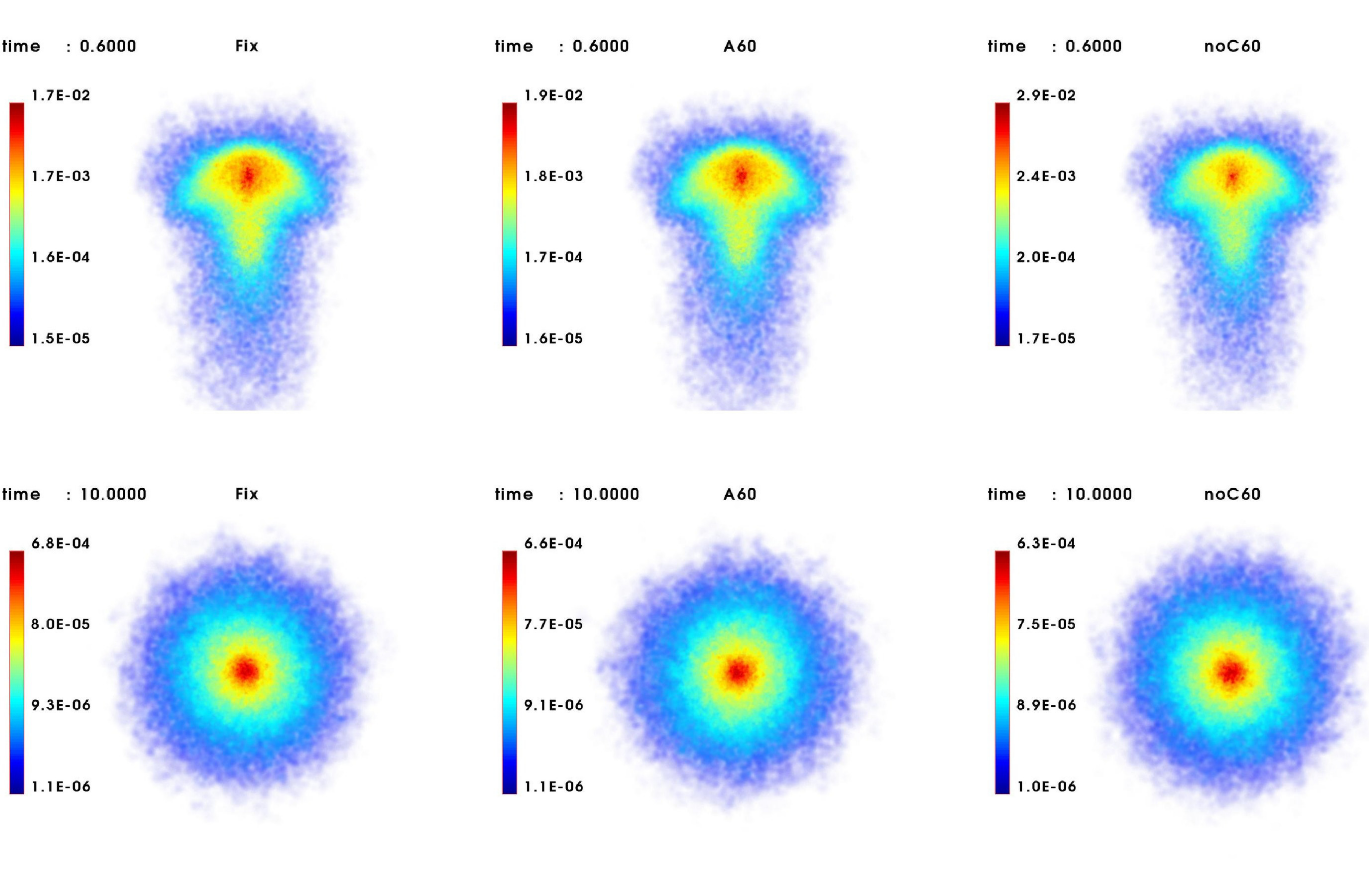}
  \caption{Inner structure of the Hernquist satellite at two representative snapshots of the merger simulation-set. The first row shows a moment where, just after the first collision with the disc, the object undergoes a phase of compression. The second row describes the situation at the end of the simulation, when the satellite has not been evolving significantly in several gigayears. The particles are coloured based on their associated density (in units of $[10^{10}\;\rm{M_{\odot}} / \rm{kpc}^3]$), whose values vary within the range indicated by the colour bar. Note that these span different intervals for the various simulations and that a given colour will in general not correspond to the same densities in all plots. The physical size displayed is $50\;\rm{kpc}$ in the upper panels and $90\;\rm{kpc}$ in the lower ones. Each column shows the results of a different softening approach, as explained in Sec.~\ref{sec:simsres}.}
  \label{merger_intruder}
\end{figure*}

\begin{figure*}
\includegraphics[width=178mm]{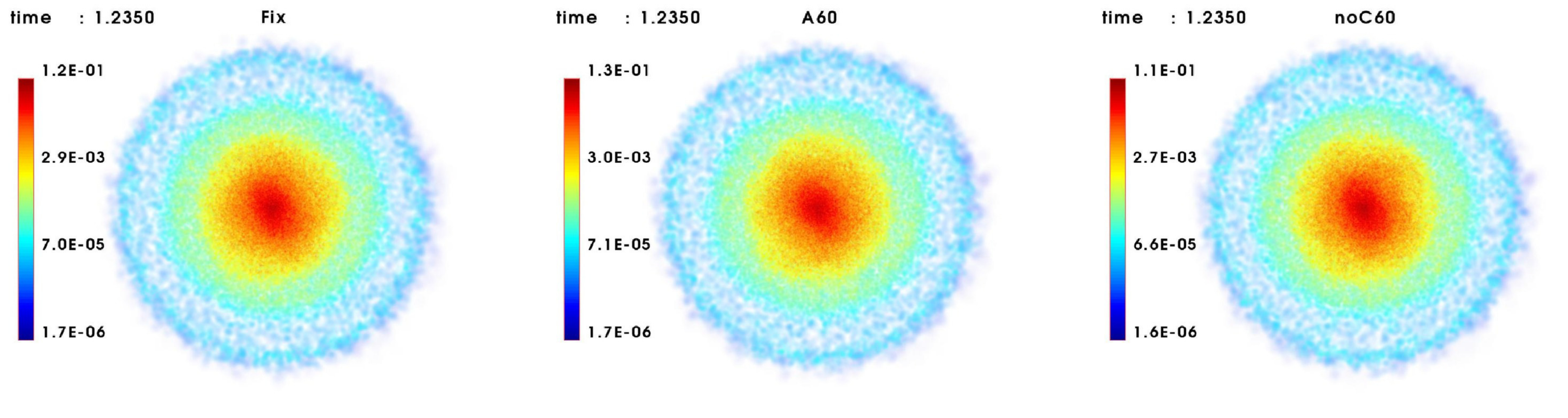}
  \caption{Face-on view of the disc component in the merger simulation-set and at the time of the ring formation. The particles are coloured based on their associated density (in units of $[10^{10}\;\rm{M_{\odot}} / \rm{kpc}^3]$), whose values vary within the range indicated by the colour bar. Note that these span different intervals for the various simulations and that a given colour will in general not correspond to the same densities in all plots. The physical size displayed is $40\;\rm{kpc}$ in each direction and for all cases. The plots show the results of different softening approaches, as explained in Sec.~\ref{sec:simsres}.}
  \label{merger_ring}
\end{figure*}

\begin{figure*}
\includegraphics[]{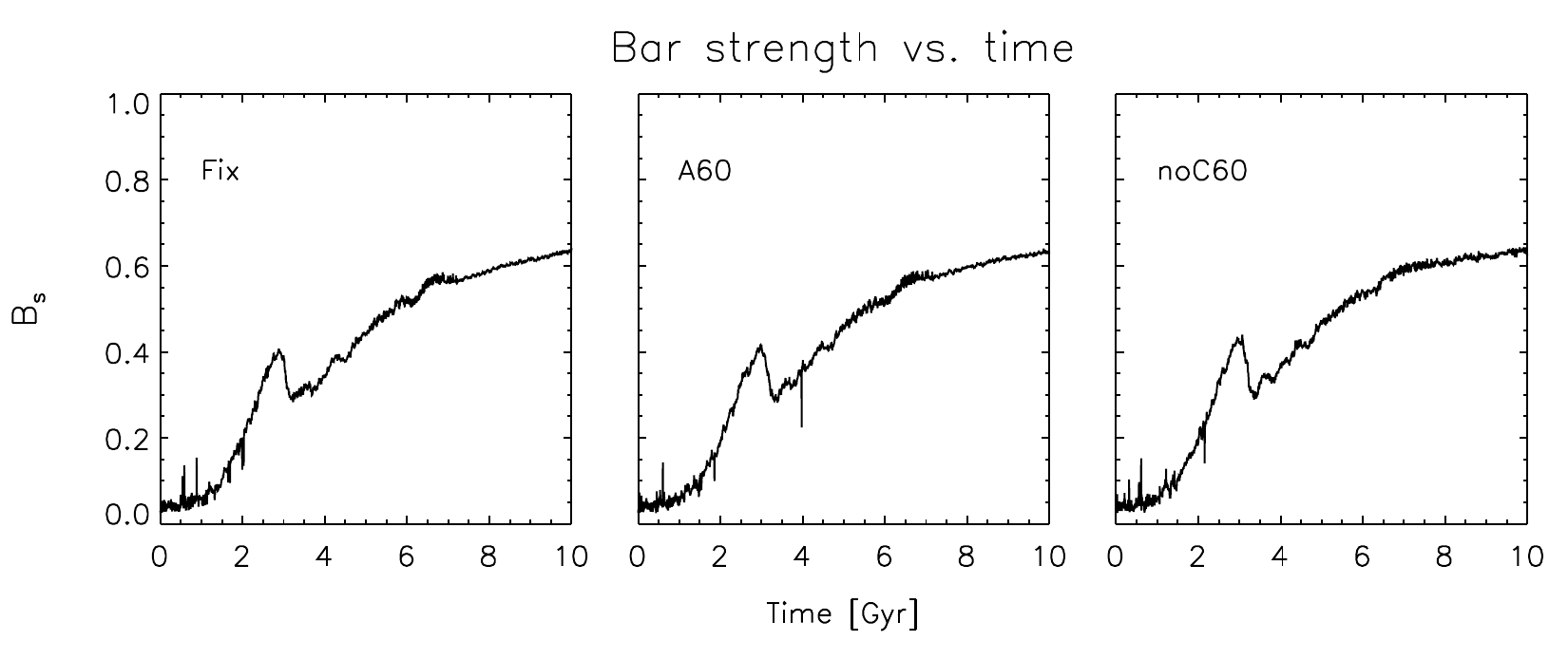}
  \caption{Bar strength as a function of time for the merger simulation-set. As for the ``cusp'' case, the bar strength is defined as the relative amplitude of the $m=2$ Fourier components of the disc (see Eq.~\ref{bs} and the explanation in Sec.~\ref{cusp}). Each plot shows the results of a different softening approach, as explained in Sec.~\ref{sec:simsres}.}
  \label{merger_bs}
\end{figure*}

We also studied a system consisting of a Hernquist sphere colliding with a galaxy-like object made of a cored halo and an exponential stellar disc. The Hernquist satellite has a mass of  $6 \times 10^{10}\;\rm{M_{\odot}}$ and a scale radius $a=9\;\rm{kpc}$. Following \cite{springeldimatteo05}, this choice results in a mass distribution comparable to that of a NFW halo of the same virial mass and concentration $c=20$. This value exceeds the median of the concentrations expected from halos in the same mass range  within a $\Lambda CDM$ context \citep[see, e.g.,][]{munozcuartas11}. However, the system has not been set up to represent a strictly realistic merger scenario; adopting an excess central concentration served the purpose of amplifying the differences between the softening approaches.
The properties of the galaxy-like object are thoroughly described e.g. in \cite{athanassoula03}. The halo has a core radius $\gamma=1.5\;\rm{kpc}$  and a mass of $2.5 \times 10^{11}\;\rm{M_{\odot}}$. The stellar disc is characterised by a scale length $h=3 \;\rm{kpc}$ and a scale height $ z_0=0.6 \;\rm{kpc}$, while its mass amounts to $5 \times 10^{10}\;\rm{M_{\odot}}$. All particles in the simulation have the same mass of $m=2.5 \times 10^5\;\rm{M_{\odot}}$; this implies a total of $1.44$ million particles - one million for the cored halo, $200\;000$ for the stellar disc and $240\;000$ for the Hernquist satellite. \\
This $1:5$  merger  occurs along a trajectory perpendicular to the plane of the disc and crossing its centre. The initial separation between the two objects is $70\;\rm{kpc}$ and the relative velocity is zero. A first passage of the satellite through the disc occurs at $t \approx 0.5\;\rm{Gyr}$ and is followed by a second collision  at $t \approx 0.8\;\rm{Gyr}$, at which time the central -- densest -- region of the Hernquist sphere decouples from its outer envelopes, reverts trajectory and crosses the disc in the opposite direction. Starting from $t \approx 2\;\rm{Gyr}$ and until the end of the simulation ($t=10\;\rm{Gyr}$), the system settles in a configuration whereby the remnant of the original satellite surrounds the central regions of the disc in the form of a dispersed halo. As far as the disc component is concerned, the geometry of the impact causes the formation of two consecutive ring-structures \cite[for the theory behind the formation of these features see, e.g.,][]{lynds76, theys76, theys77, 
toomre78}, at 
$t \approx 0.9\;\rm{Gyr}$ and $t \approx 1.2\;\rm{Gyr}$ respectively. Eventually, a bar develops and follows its traditional evolutionary pattern.\\
We have simulated the system with three of the softening approaches introduced at the beginning of Sec.~\ref{sec:simsres}: fixed softening (Fix), adaptive softening with correction term (A60) and adaptive softening without correction term (noC60). In Fix, $\epsilon$ was again chosen to be $0.05\;\rm{kpc}$. This was the fiducial value used in previous simulations of the halo+disc system and, as far as the Hernquist sphere is concerned, it corresponds to $\approx 1/5$ of the mean interparticle separation within $a$. For the simulations with adaptive softening, we adopted the fiducial choice $N_{ngbs}=60\pm0.3$. Foreseeing poor conservation properties in a dynamically more challenging scenario, we have not performed a run with $N_{ngbs}=30$ as for the ``cusp'' set. For all the simulations performed (included, notably, noC60) energy conservation is guaranteed at a $0.2\%$ level. \\
The evolution of the system in the different simulations is very similar. By inspecting density maps of the Hernquist component, we find that the adaptive runs provide the highest registered densities during the most violent events and in the region of highest compressions (see, e.g., the upper plots of Fig.~\ref{merger_intruder}). In less extreme phases, the induced matter density is very similar to Fix and sometimes it peaks to slightly lower values; in these configurations, the run noC60 tends to provide lower densities than A60 (see the plots at the bottom of Fig.~\ref{merger_intruder}).\\
Similar conclusions hold for the evolution of the disc component. As the density maps in Fig.~\ref{merger_ring} show, the ring features are equally-well resolved in the three runs. In the snapshot shown ($t=1.235\;\rm{Gyr}$) the density peaks at a $10\%$-higher value in A60 with respect to Fix (and $\approx 20\%$-higher with respect to noC60), but the relative differences between the three cases may slightly vary from one time to the other. The evolution of the bar strength with time is also remarkably similar in the three simulations\footnote{The local peak in $B_s$ at the end of the first phase of bar growth is $\approx 25\%$ lower than it would be in the absence of a merger. Other than for this, the evolution of the bar strength with time is remarkably similar to what it would be if the galaxy was isolated. The differences with the behaviour obtained in the ``cusp'' simulations (see Fig.~\ref{nfw_bs_and_ps}) is to be ascribed to the different halo properties in the two cases \citep[i.e. core vs. cusp, see 
][for a review]{athanassoula13}}, as can be seen in Fig.~\ref{merger_bs}.\\
Finally, no significant differences have been found in the representation and evolution of the halo component, whose profile and central density match among the different runs and throughout the simulation. \\

\section{Summary and Conclusions}\label{concl}
In this study we have explored the effect of different softening approaches on N-body simulations of idealised galaxies. Alongside the standard method involving a softening length fixed in space and time, we performed runs where this scale was allowed to vary from particle to particle and throughout the simulation. The underlying algorithm, developed by \cite{pm07} and implemented in \gadgt$\,$ by \cite{iannuzzi11}, provides, in fact, an adaptive gravitational resolution following the features of the density field in the system (see Fig.~\ref{agsh_combined}). 
The systems under consideration here are an isolated galaxy-like object and an idealised merger configuration. Other than for the low-mass player in the merger event, which is modelled as a one-component Hernquist sphere, the other systems consist of equilibrium disc+halo configurations. The halos considered have either cores or cusps, while the stellar disc has always a standard exponential profile.\\
Such idealised simulations -- also termed as ``dynamical'' -- where the system is let to evolve passively in isolation, are generally performed to study the internally-driven dynamical evolution of disc galaxies. Adopting few millions of particles to model an object the mass of the Milky Way, they can afford a gravitational resolution of a few tens of parsecs. By simulating these systems with adaptive softening, thereby increasing the resolution in the over-dense regions and lowering it otherwise, we were interested in assessing whether or not the results from standard simulations with fixed softening are hampered by a resolution limit.\\
In Sec.~\ref{cusp} we showed the results of a ten-gigayear evolution of one such galaxy-like object left in isolation. The results of the different runs, as far as the disc component is concerned, are compatible. The resulting bar and its temporal evolution show striking similarities regardless of the adopted softening approach (Fig.~\ref{nfw_isod}, \ref{nfw_dens_prof} and \ref{nfw_bs_and_ps}). Even in the simulation where $\approx 40\%$ of the disc particles had associated softening smaller (up to a factor of eight) than the fiducial value used in the standard simulation, the bar strength and central density did not show significant differences (at least not in the regime where this specific adaptive simulation was fully reliable).\\
In Sec.~\ref{merger} we report on the results of a simple $1:5$ merger simulation, where a Hernquist satellite crossed the centre of a galaxy-system on a trajectory perpendicular to the plane of the stellar disc. The merger event unfolds in a very similar fashion in the three simulations performed. The rings in the disc component develop at the same time and with broadly the same features (Fig.~\ref{merger_ring}). At later times, when a bar finally develops, its growth is found to be remarkably similar, even quantitatively, regardless of the adopted softening approach (Fig.~\ref{merger_bs}). Generally, in moments of maximum compression, the runs with adaptive softening tend to provide higher densities in the particle distribution. During quieter phases the situation may vary from snapshot to snapshot, with fixed softening and fully-conservative adaptive softening providing, typically, the highest registered densities (Fig.~\ref{merger_intruder}). \\
As far as the dynamical evolution of the disc is concerned, we can therefore conclude that adopting a fixed value for the softening length everywhere in the system, even if this means losing resolution in the innermost regions, is not significantly affecting the results. Presumably, in the systems under consideration it is still possible to identify a value for $\epsilon$ that stands as an optimal compromise between the resolution needed in the innermost and outermost regions so that the system is overall faithfully modelled. In a dynamically more challenging scenario, as in the merger simulation presented, the differences between a fixed and an adaptive resolution show up in the most violent phases involving high density contrasts. No trace of these differences is inherited by the temporal evolution of the disc component, though, and also the late-time properties of the satellite are very similar in all the runs. Even in this case, then, adopting an appropriate fixed value for the softening length turns out 
to be a choice accurate enough to grasp the main results from the simulated system.

\section*{Acknowledgments}
We acknowledge financial support to the DAGAL network from the People Programme (Marie Curie Actions)
of the European Union's Seventh Framework Programme FP7/2007- 2013/ under REA grant agreement number
PITN-GA-2011-289313. We thank Albert Bosma, Klaus Dolag and Sergey Rodionov for useful discussions and comments.
For Fig.~\ref{agsh_combined},~\ref{merger_intruder} and~\ref{merger_ring} we made use of the UNS and GLNEMO software (\url{http://projets.lam.fr/projects/uns_projects}, \url{http://projets.lam.fr/projects/glnemo2}).

\bibliography{biblio}{}
\bibliographystyle{mn2e}

\appendix
\section[]{The choice of $\epsilon$}
\label{app}

\begin{figure}
\includegraphics[width=84mm]{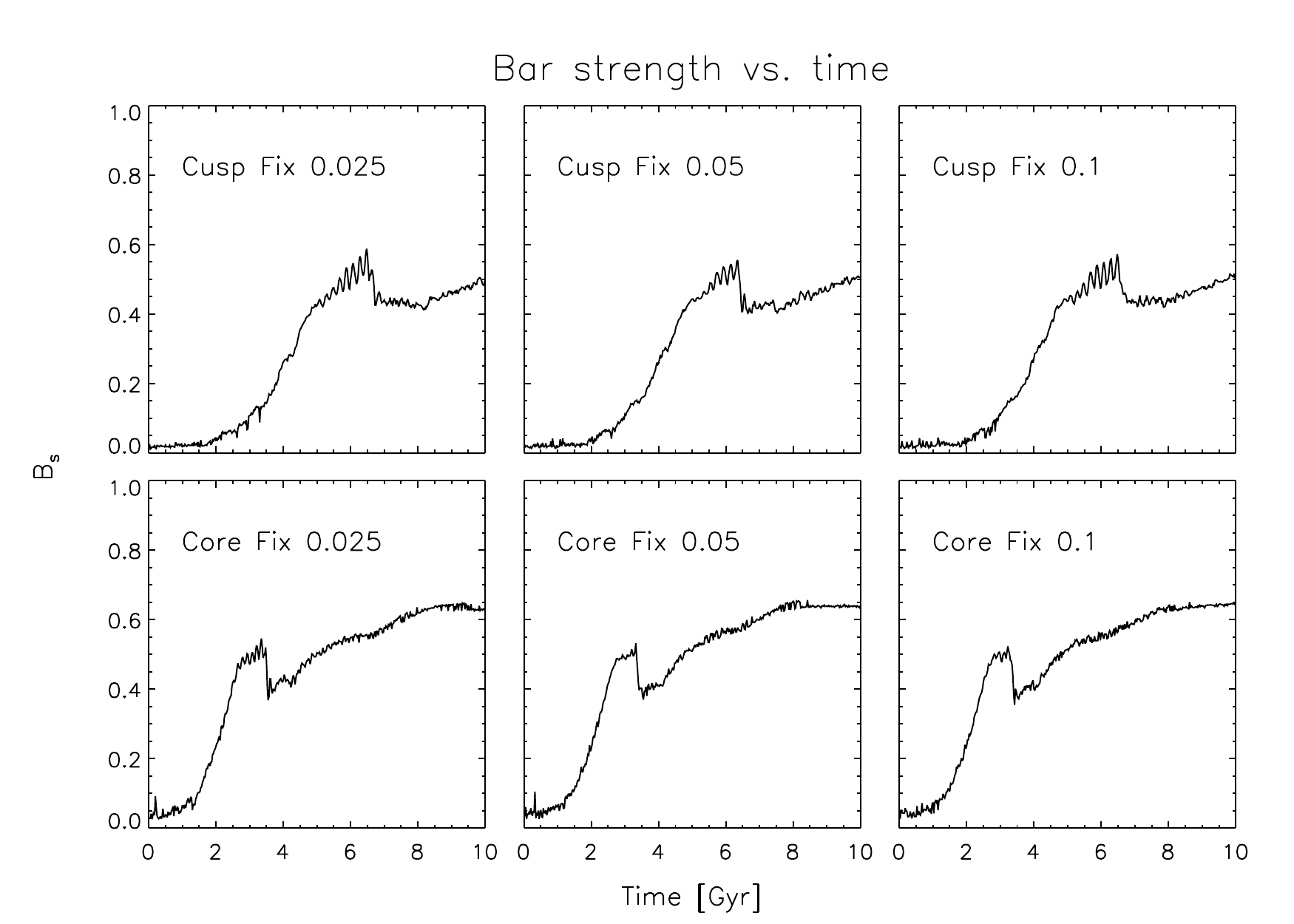}
  \caption{Bar strength as a function of time for the ``cusp'' and ``core'' simulations (top and bottom row, respectively). 
  The bar strength is defined as the relative amplitude of the $m=2$ Fourier components of the disc (see Eq.~\ref{bs} and the explanation in Sec.~\ref{cusp}). Each column corresponds to a different adopted $\epsilon$ - $0.025, 0.05$ and $0.1\;\rm{kpc}$ from left to right, respectively.}
  \label{bs_app}
\end{figure}

\begin{figure}
\includegraphics[width=84mm]{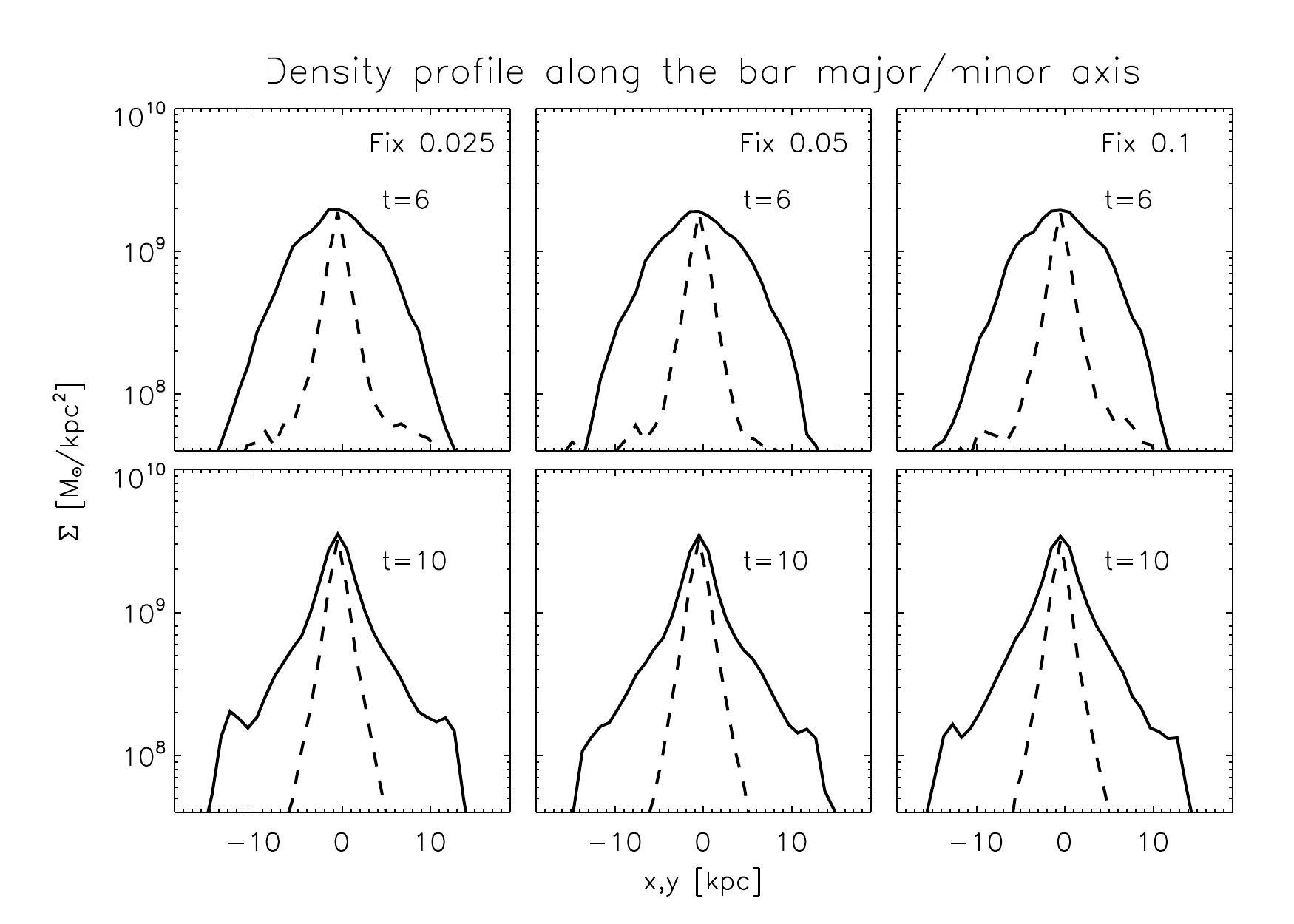}
  \caption{Projected density profiles along the bar major and minor axes (solid and dashed lines, respectively) for the ``cusp'' simulations. The surface density is computed by projecting the particles on to the plane of the disc and by selecting those within $0.1\;\rm{kpc}$ from the axis under consideration. Results are shown at $t=6\;\rm{Gyr}$, approximately at the end of the initial growth-phase for the bar, and at $t=10\;\rm{Gyr}$, marking the end of the simulation. Each column corresponds to a different adopted $\epsilon$ - $0.025, 0.05$ and $0.1\;\rm{kpc}$ from left to right, respectively.}
  \label{core_dens_prof_app}
\end{figure}

\begin{figure}
\includegraphics[width=84mm]{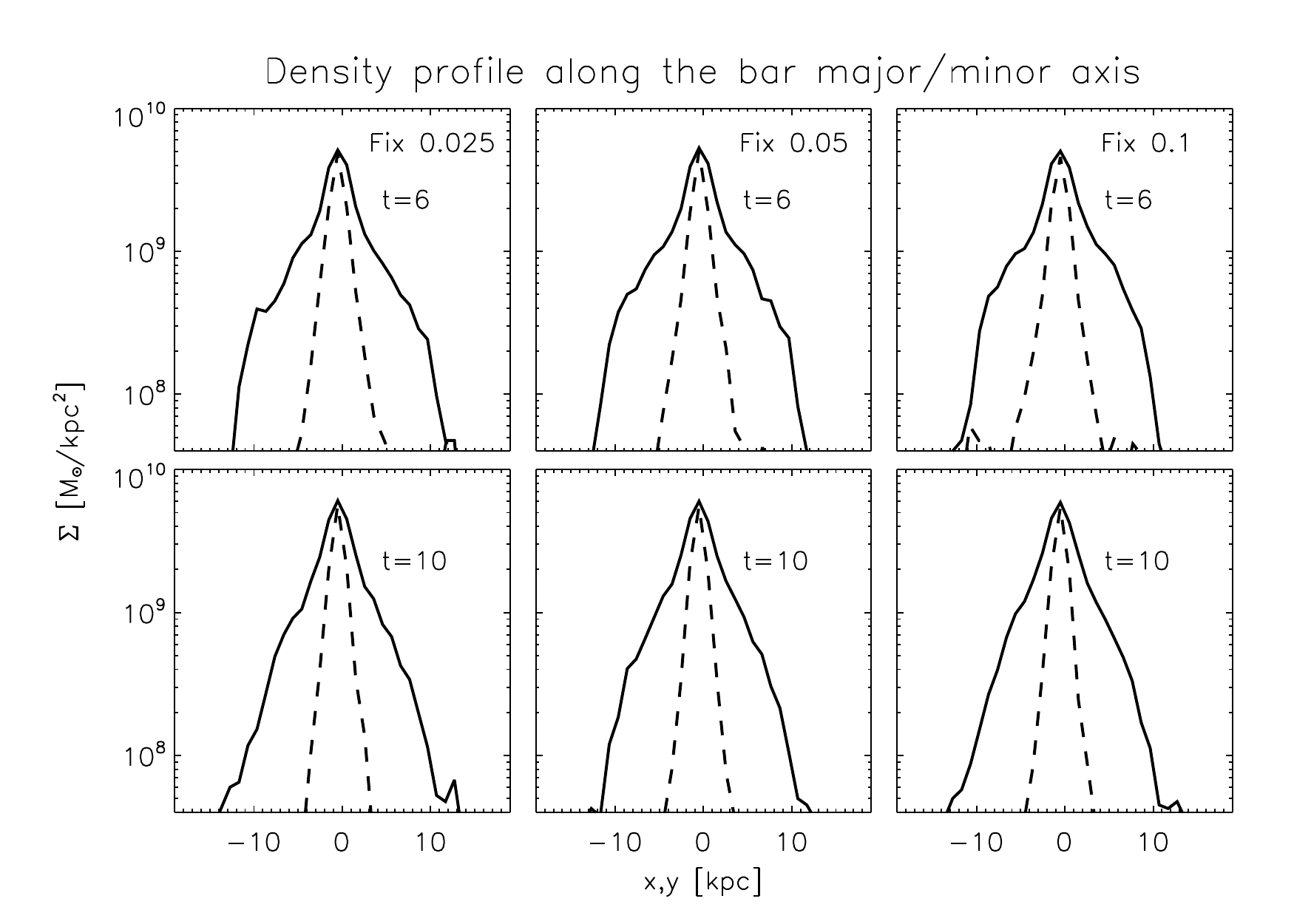}
  \caption{As for Fig.~\ref{core_dens_prof_app}, but for the ``core'' case.}
  \label{cusp_dens_prof_app}
\end{figure}

In this paper, we adopted $\epsilon = 0.05\;\rm{kpc}$ for all simulations performed with fixed softening. As explained in the text, this is the fiducial valued adopted for identical and/or similar configurations in previous works. We did, though, run both the isolated ``cusp'' and ``core'' models with different values of $\epsilon$ in order to further assess the meaningfulness of our choice. For completeness, we report here the results of our analysis.\\
Fig.~\ref{bs_app} shows the behaviour of the bar strength ($B_s$) as a function of time for the ``cusp'' and ``core'' model (top and bottom row, respectively) when $\epsilon$ is set to $0.025, 0.05$ and $0.1\;\rm{kpc}$ (left, middle and right column, respectively). No significant difference is found as $\epsilon$ is halved or doubled with respect to its fiducial value. A similar conclusion holds for the projected density profile along the major/minor axis of the bar, plotted for the same models in Fig.~\ref{core_dens_prof_app} and \ref{cusp_dens_prof_app}. As $\epsilon$ varies, differences at the level of a few per cent or less are found from one bin to the other, but overall the profiles stay remarkably similar. As one can imagine at this point, neither the velocity curves nor the isodensity contours show appreciable variations with $\epsilon$ in the considered range and we do not show the relative plots for brevity. \\
One would naively expect the model with $\epsilon = 0.025\;\rm{kpc}$ to provide an enhanced concentration of particles and the opposite for the model with $\epsilon = 0.1\;\rm{kpc}$. Lacking a macroscopic evidence for this hypothesis and its consequences, we monitored the time variation in the number of particles (halo and disc component together) enclosed within spheres of radii $r = 0.15, 0.2, 0.3, 0.5\;\rm{kpc}$ for both the ``cusp'' and ``core'' models with different $\epsilon$. Softening-induced differences appear at $r \leq 0.2\;\rm{kpc}$, but not always as expected. Indeed, in the ``core'' model a small softening does induce a progressively (up to 25\%) larger concentration of particles at small radii, but the opposite is true for the ``cusp'' model. However, in both cases these differences manifest themselves at radii containing only $\approx 100,\;300$ particles respectively. Overall, we do not find conclusive evidence that adopting $\epsilon = 0.025\;\rm{kpc}$ would be advantageous for our 
simulations.\\
\begin{figure}
\includegraphics[width=84mm]{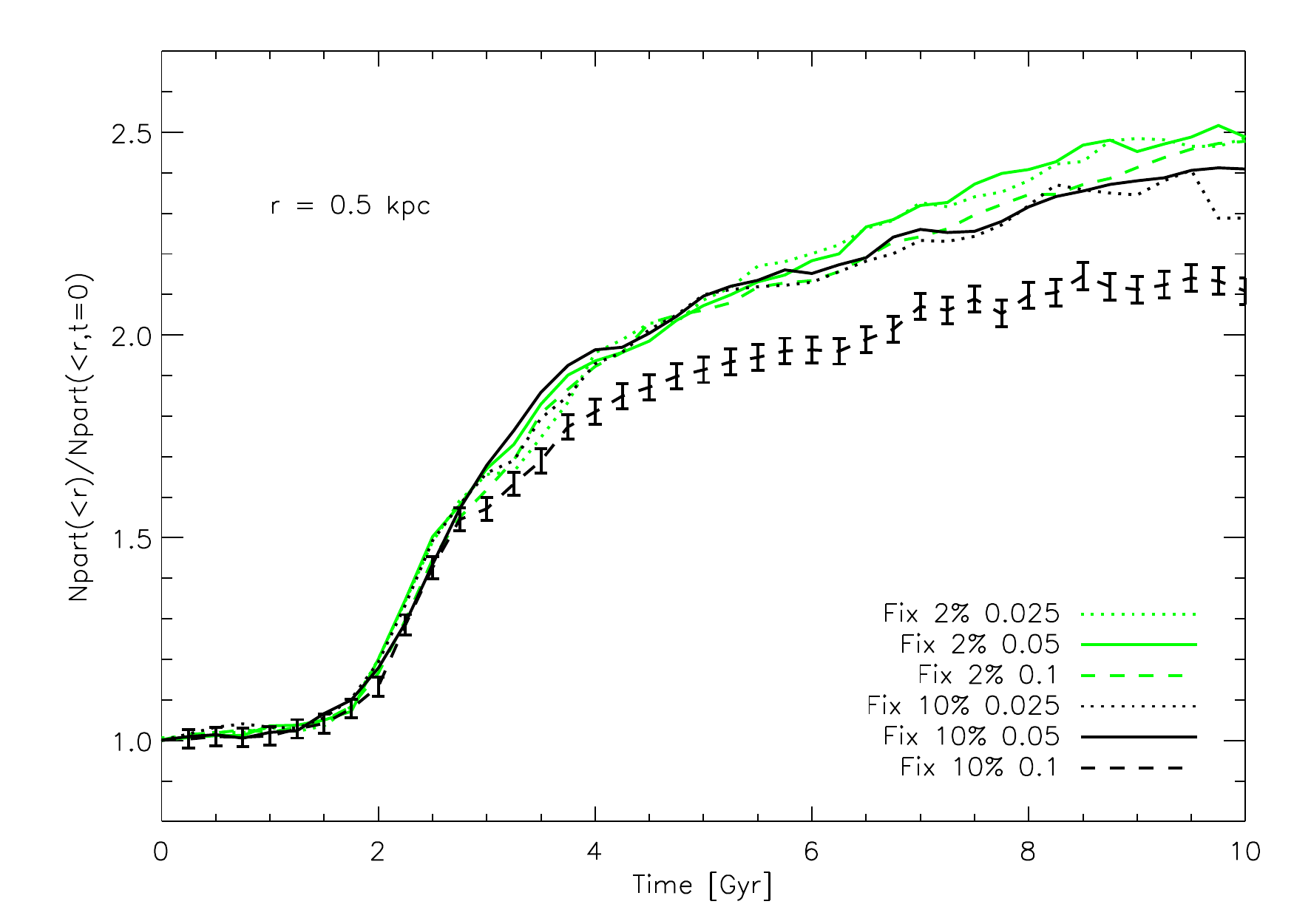}
  \caption{Time evolution of the number of particles enclosed in a sphere or radius $r = 0.5\;\rm{kpc}$ centred on the centre of mass of the system, normalised to the $t=0$ value. The analysis is done on the ``core'' simulation and particles of both ``disc'' and ``halo'' type are considered. Two sets of runs are shown: for the green curves, the maximum allowed timestep was set to $\approx 2\%$ of the dynamical time, while for the black curves to $\approx 10\%$ of the dynamical time. Within each set, the solid curve corresponds to the fiducial choice $\epsilon = 0.05\;\rm{kpc}$, the dotted to $\epsilon = 0.025\;\rm{kpc}$ and the dashed to $\epsilon = 0.1\;\rm{kpc}$. The error bars overplotted on the black-dashed curve mark the amplitude of $\sqrt{N}$, for reference.}
  \label{r0.5_app}
\end{figure}
As far as the run with the largest softening is concerned, we register a very similar behaviour to the fiducial one even in this last test. We checked that this is due to our somewhat conservative choice for the maximum timestep a particle is allowed to adopt at any time during the simulation. We set this value to $\approx2\%$ of the dynamical time and this effectively causes the two simulations to have the same resulting time-integration accuracy, even though the softening varies by a factor of two. If we relax the limit on the maximum timestep and set it to $\approx 10\%$ of the dynamical time, thereby obtaining two simulations with different assigned timesteps, we recover the expected result in terms of larger softening inducing lower central concentrations (an effect visible already at $r = 0.5\;\rm{kpc}$, see Fig.~\ref{r0.5_app}). We note that this behaviour cannot be merely attributed to a larger softening and that it is, in fact, induced by a sloppier time integration in combination with a larger 
softening. This recovered difference in the number of enclosed particles at small radii does not result in appreciable changes in the behaviour of $B_s$ in time, which remains compatible with the results from the other runs with smaller softenings, as well as those with maximum timestep  $\approx2\%$ of the dynamical time. The projected density distribution changes as expected: the profile now peaks at $5-10\%$ lower values in the inner $0.5\;\rm{kpc}$. \\
We note that even with this latter choice of maximum timestep the simulation with the smallest softenings does not induce an enhanced central concentration of particles with respect to the run using the fiducial value for $\epsilon$ (see the solid and dotted black lines in Fig.~\ref{r0.5_app}), nor do these two differ in the macroscopic quantities examined. This confirms that no advantage would be brought about by adopting $\epsilon = 0.025\;\rm{kpc}$ in the configuration of interest here.

\label{lastpage}

\end{document}